\documentclass[twocolumn,preprintnumbers,amsmath,amssymb]{revtex4}
\usepackage{graphicx}
\usepackage{color}
\usepackage{dcolumn}
\usepackage{bm}
\usepackage[latin1]{inputenc} 
\usepackage[normalem]{ulem}

\newcommand{\comment}[1]{}

\begin{document}
\setlength{\unitlength}{0.7\textwidth} \preprint{}

\title{Dissipation in unsteady turbulence}


\author{Wouter J.T. Bos$^1$}
\author{Robert Rubinstein$^2$}
\affiliation{$^1$ LMFA, CNRS, Ecole Centrale de Lyon -
Universit\'e de Lyon -  Ecully, France\\
$^2$ 
Newport News, VA, USA}





\begin{abstract}
Recent experiments and simulations have shown that unsteady turbulent flows, before reaching a dynamic equilibrium state, display a universal behaviour. We show that the observed universal non-equilibrium scaling can be explained using a non-equilibrium correction of Kolmogorov's energy spectrum.
Given the universality of the experimental and numerical observations, the ideas presented here lay the foundation for the modeling of a wide class of unsteady turbulent flows.
\end{abstract}

\maketitle

Turbulent flows are characterized by spatio-temporal fluctuations on a wide range of scales. The general challenge of turbulence research is to find a universal description of the statistics of these fluctuations. The derivation of such a description from the governing equations has proven extremely difficult \cite{Leslie,FrischBook,KraichnanDIA,Kraichnan65}, but experiments and simulations show that several features are fairly universal. Two of such features, and corner-stones of the description of turbulent flows, are Taylor's dissipation rate estimate \cite{Taylor1935}, and Kolmogorov's inertial range description \cite{Kolmogorov}. In this letter we will show how these two features are related in unsteady turbulence. In particular, we will show that a number of features in recent experiments and numerical simulations can be explained using the unsteady analog of Kolmogorov's inertial range spectrum.

Kolmogorov's concepts, introduced in the 1940s, state that the energy distribution among scales at sufficiently high Reynolds number is completely determined by the scale-size and the energy flux through scales, for scale-sizes sufficiently small compared to the most energetic eddies, and sufficiently large compared to the smallest, dissipative scales. In this range the energy spectrum is approximately given by the relation,
\begin{equation}\label{eq:K41}
 E(\kappa,t)=C_K \epsilon(t)^{2/3}\kappa^{-5/3},
\end{equation}
where $\epsilon(t)$ is the average dissipation rate, $\kappa$ the wavenumber, and $C_K\approx 1.5$ a constant. This relation is observed, to a good approximation, in a wide range of turbulent flows. Taylor's dissipation rate estimate, 
\begin{equation}\label{eq:Ceps}
 \epsilon(t)=C_\epsilon \frac{U(t)^3}{L(t)},
\end{equation}
relates the dissipation rate, which is in principle a small scale quantity, to the dynamics of the large-scale quantities  $U$, the RMS velocity and $L$ the integral lengthscale at high Reynolds numbers \cite{Sreeni84,Kaneda2003}. The insight that the dissipation can be modeled using large-scale quantities, allows the formulation of simple engineering models that need not take into account the multi-scale character of turbulence. Both relations are intimately related \cite{Lumley92,Bos2007-2} and an estimate of the constant $C_\epsilon$ can be obtained using relation (\ref{eq:K41}) (details are given below). The quantity $C_\epsilon$ can be expressed as a function of two distinct Reynolds numbers, through the relation,
\begin{equation}\label{eq:CepsRR1}
C_\epsilon \sim \frac{R_L(t)}{R_\lambda(t)^2}
\end{equation}
where
\begin{equation}\label{eq:Reynolds}
R_L(t)=\frac{U(t)L(t)}{\nu} ,~~~~~~ R_\lambda(t)=\sqrt{15\frac{U(t)^4}{\nu \epsilon(t)}},
\end{equation}
where it can be noted that $C_\epsilon$ is independent of the viscosity $\nu$. These relations are expected to hold in fully developed high Reynolds number flows. Recent experimental studies at Imperial College London considering decaying wind-tunnel turbulence behind different types of turbulence-generating grids \cite{Seoud2007,Mazellier2010,Valente2012,Valente2014} have focused on the transient period before the turbulence is fully developed and have shown that the decay obeys, to a good approximation, a universal law, 
\begin{equation}\label{eq:CepsJCV}
 C_\epsilon\sim \frac{\sqrt{R_L(0)}}{R_\lambda(t)}
\end{equation}
where $R_L(0)$ is determined by the initial conditions. Other research groups confirmed the results in independent grid-turbulence experiments \cite{Discetti2013,Hearst2014,Isaza2014} and direct numerical simulations \cite{Nagata2013}. The scaling observed in these experiments seems more general than the case of freely decaying grid-turbulence only, since experiments and simulations of the wakes generated by plates with irregular edges show the same tendency \cite{Nedic2013,Dairay2015}. Recently, it was shown that in yet another different type of turbulent flow, where the kinetic energy is maintained at a certain level through an external forcing, the fluctuations of kinetic energy and dissipation around the long-time-averaged state can be described by the same law \cite{Goto2015}.

It is noted that expression (\ref{eq:CepsJCV}) is radically different from (\ref{eq:CepsRR1}). Since, as stated before, $C_\epsilon$ can be related to Kolmogorov's energy spectrum (\ref{eq:K41}), (\ref{eq:CepsJCV}) might suggest a departure from (\ref{eq:K41}) during the transient, but this is not observed. Our analysis explains these puzzling results. In particular it is shown that the observation of (\ref{eq:CepsJCV}) is related to a sub-dominant correction to Kolmogorov's energy spectrum  first proposed in \cite{Yoshizawa1994}. 
We reproduce a simple derivation of the non-equilibrium correction here. The same results were obtained by \cite{Woodruff2006,Rubinstein2005} using similarity arguments, using Kovaznay's closure in \cite{Horiuti2013} and using more sophisticated closures  in \cite{Yoshizawa1994,Bos2007-3}. 

We start from the evolution equation for the kinetic energy spectrum at high Reynolds numbers at scales where both production and dissipation mechanisms can be neglected,
\begin{equation}
 \partial_t E(\kappa,t)=-\partial_\kappa \Pi(\kappa,t),
\end{equation}
where $\Pi(\kappa,t)$ is a flux of energy which should vanish at $\kappa=0$ and $\kappa=\infty$. We make the assumption that we  can decompose the energy spectrum into its {\it equilibrium} and {\it non-equilibrium} parts, 
\begin{equation}\label{eq:EE2}
E(\kappa,t)=\overline E(\kappa,t)+\widetilde E(\kappa,t). 
\end{equation}
It is extremely important for the following to note that both parts are a function of time and that this is not a separation of the energy distribution in a steady and an unsteady part. The equilibrium part of the turbulence corresponds to the part for which the flux is not a function of scale $\Pi(\kappa,t)= \epsilon(t)$, yielding for the evolution of the spectrum,
\begin{equation}\label{eq:Ebart}
 \partial_t \overline E(\kappa,t)=-\partial_\kappa \widetilde \Pi(\kappa,t),
\end{equation}
where we have assumed that $\widetilde E(\kappa,t)\ll \overline E(\kappa,t)$. It is at this point that we need the introduction of an assumption on the functional form of the flux. In the present derivation, considering Kovaznay's model \cite{Kovaznay},
\begin{equation}\label{eq:Kov}
 \Pi(\kappa,t)=C_K^{-3/2} \kappa^{5/2} E(\kappa,t)^{3/2},
\end{equation}
immediately yields, when $\Pi(\kappa,t)=\epsilon(t)$, that $\overline E(\kappa,t)$ is given by, 
\begin{equation}\label{eq:K41bar}
 \overline E(\kappa,t)= C_K \epsilon(t)^{2/3}\kappa^{-5/3}.
\end{equation}
Introducing (\ref{eq:EE2}) into (\ref{eq:Kov}) yields for small perturbations,
\begin{equation}\label{eq:123}
 \Pi(\kappa,t)=C_K^{-3/2} \kappa^{5/2} \overline E(\kappa,t)^{3/2}\left(1+\frac{3}{2}\frac{\widetilde E(\kappa,t)}{\overline E(\kappa,t)} \right),
\end{equation}
substituting this into expression (\ref{eq:Ebart}) gives upon integration
\begin{equation}\label{eq:Y94}
 \widetilde E(\kappa,t)=C_K \Omega_\epsilon(t)\epsilon(t)^{1/3}\kappa^{-7/3},
\end{equation}
with
\begin{equation}
\Omega_\epsilon(t)=\frac{2C_K}{3}\frac{\dot \epsilon(t)}{\epsilon(t)}. 
\end{equation}
It is this new frequency $\Omega_\epsilon$ in the dynamics which allows to find the $k^{-7/3}$ scaling in (\ref{eq:Y94}) as a first linear correction to classical scaling, as for the shear-stress spectrum in homogeneous shearflow, where the mean-velocity gradient is introduced as typical frequency \cite{Lumley1967}.

Both spectra (\ref{eq:K41bar}) and (\ref{eq:Y94}) are a function of time. The equilibrium part describes thus not necessarily a steady state, and temporal fluctuations are therefore not purely described by (\ref{eq:Y94}), since if they are slow enough, they will have time to adapt to the equilibrium distribution (\ref{eq:K41bar}). 
The observation of the non-equilibrium scaling (\ref{eq:Y94}) is not straightforward, since it is subdominant with respect to the Kolmogorov spectrum (\ref{eq:K41bar}). Conditional averaging allows however to extract the unsteady energy spectrum, as was illustrated in reference [\onlinecite{Horiuti2013}], where a clear $\kappa^{-7/3}$ wavenumber spectrum was observed in a statistically steady turbulent flow simulation. The difference in wavenumber scaling between $\overline E(\kappa,t)$ and $\widetilde E(\kappa,t)$ is the origin of the observations of a non-classical, but universal transient scaling of the dissipation rate. We will elaborate on that in the following. We note that the possible relevance of the spectrum suggested by Yoshizawa (\ref{eq:Y94}) to estimate the dissipation rate in unsteady turbulence was already mentioned in reference \cite{Goto2016}.

To simplify the considerations we assume the two scalings (\ref{eq:K41bar}) and (\ref{eq:Y94}) to hold in the wavenumber interval between $\kappa_0$, the forcing scale and $\kappa_\eta$, the Kolmogorov scale, given by
\begin{equation}
 \kappa_\eta\sim \frac{\overline \epsilon(t)^{1/4}}{\nu^{3/4}}.
\end{equation}
Outside this interval the kinetic energy is assumed to be zero, for analytical convenience. 
We have also considered more complicated spectra adding a more realistic infrared energy range (as in \cite{Bos2007-2}) and the results hereafter were shown to be robust.  
All quantities will be decomposed into their equilibrium part, indicated by a bar and their non-equilibrium part, indicated by a tilde. For instance $C_\epsilon=\overline C_\epsilon+\widetilde C_\epsilon$, $\epsilon=\overline \epsilon+\widetilde \epsilon$, etc. The equilibrium kinetic energy is computed by integrating expression (\ref{eq:K41bar}),
\begin{equation}\label{eq:intE}
\overline k(t)=\int \overline E(\kappa,t) d\kappa=(3/2)\overline {U}(t)^2,  
\end{equation}
and, similarly the non-equilibrium energy is obtained from equation (\ref{eq:Y94}). 
The integral lengthscale is defined by
\begin{equation}\label{eq:intL}
 L(t)=\frac{3\pi}{4}\frac{\int \kappa^{-1} E(\kappa,t) d\kappa}{\int E(\kappa,t) d\kappa}\equiv \frac{3\pi}{4}\frac{\mathcal I(t)}{k(t)}.
\end{equation}
where we introduced $\mathcal I(t)\equiv \int \kappa^{-1} E(\kappa,t) d\kappa$ for later convenience. The dissipation can be computed from the energy spectrum by 
\begin{equation}\label{eq:inteps}
\epsilon(t)=2\nu\int \kappa^{2} E(\kappa,t) d\kappa,
\end{equation}
where all integrals are evaluated on the interval $[\kappa_0,\kappa_\eta]$. 
Carrying out these integrals using (\ref{eq:K41bar}) to evaluate $\overline U$ and $\overline L$ (expressions (\ref{eq:intE}) and (\ref{eq:intL})),  and substituting these expressions in Kolmogorov's and Taylor's expressions (\ref{eq:K41}) and (\ref{eq:Ceps}), it is immediately found that the equilibrium value of the normalized dissipation rate is 
\begin{equation}\label{eq:CepsBarNum}
\overline C_\epsilon=\frac{3\pi}{10}C_K^{-3/2}\approx 0.51. 
\end{equation}
This value is thus the inertial range estimate of $C_\epsilon$, assuming a spectrum given by (\ref{eq:K41}) on the interval $[\kappa_0,\kappa_\eta]$. Despite such gross assumptions on the shape of the energy spectrum, its value is actually close to the value observed in direct numerical simulations of forced high Reynolds numbers turbulence where values around $0.5$ are observed \cite{Kaneda2003}. In the following we will omit the time-dependence of the different quantities to lighten the notation. It should however be kept in mind, as we stressed before, that both the equilibrium and the non-equilibrium quantities can depend on time.

Since $C_\epsilon\sim \epsilon \mathcal I/k^{5/2}$, we can write without any approximations 
\begin{equation}\label{eq:Cepsnoapp}
\frac{C_\epsilon}{\overline C_\epsilon}=\frac{\left(1+\frac{\widetilde \epsilon}{\overline \epsilon}\right)\left(1+\frac{\widetilde {\mathcal I}}{\overline {\mathcal I}}\right)}{\left(1+\frac{\widetilde k}{\overline k}\right)^{5/2}}.
\end{equation}
The different quantities in this expression are obtained by integrating the expressions (\ref{eq:intE}), (\ref{eq:intL}) and (\ref{eq:inteps}) over the interval $[\kappa_0,\kappa_\eta]$, using the spectra (\ref{eq:K41bar}) and (\ref{eq:Y94}) for the equilibrium  and nonequilibrium contributions, respectively. For instance, it is found that 
\begin{equation}\label{eq:koverk}
 \frac{\widetilde \epsilon}{\overline \epsilon}=\frac{2\Omega_\epsilon}{\epsilon^{1/3}\kappa_\eta^{2/3}} \textrm{~~and~~}    \frac{\widetilde k}{\overline k}=\frac{\Omega_\epsilon}{2\epsilon^{1/3}\kappa_0^{2/3}},
\end{equation}
where we have assumed $\kappa_0\ll \kappa_\eta$. Since in the equilibrium state $\kappa_0/\kappa_\eta\sim R_\lambda^{-3/2}$, we find that 
\begin{equation}
\frac{\widetilde \epsilon}{\overline \epsilon} \sim R_\lambda^{-1}\frac{\widetilde k}{\overline k}, 
\end{equation}
which is a direct consequence of the $k^{-7/3}$ scaling of $\widetilde E(k)$. This shows that the $\widetilde \epsilon/\overline \epsilon$ term in (\ref{eq:Cepsnoapp}) is negligible. This indicates also that the temporal dissipation rate fluctuations observed in \cite{Goto2015,Goto2016} are mainly related to the equilibrium distribution $\overline E(k,t)$ of the flow and negligibly contribute to the non-equilibrium part of the dissipation rate. We further find that $\widetilde {\mathcal I}/\overline {\mathcal I}=(10/7) \widetilde k/\overline k$. The expression for large Reynolds numbers is therefore
\begin{equation}\label{eq:Ceps52}
\frac{C_\epsilon}{\overline C_\epsilon}\approx \frac{\left(1+\frac{10}{7}\frac{\widetilde k}{\overline k}\right)}{\left(1+\frac{\widetilde k}{\overline k}\right)^{5/2}}.
\end{equation}
Evaluating the Reynolds number one finds analogously,  
\begin{equation}
\frac{R_\lambda}{\overline R_\lambda}\approx \left(1+\frac{\widetilde k}{\overline k}\right).
\end{equation}
where $\overline R_\lambda$ is given by (\ref{eq:Reynolds}) using the equilibrium values $\overline U$ and $\overline \epsilon$.
To obtain these two expressions we have thus only assumed that the Reynolds number is high and that the energy spectrum can be represented by (\ref{eq:K41bar}) and (\ref{eq:Y94}) between $\kappa_0$ and $\kappa_\eta$.  We consider the case where $\widetilde k/\overline k$ is small, for which the non-equilibrium scaling (\ref{eq:Y94}) was derived, so that we can use a Taylor-expansion to rewrite (\ref{eq:Ceps52}) as 
\begin{equation}\label{eq:Ceps1514}
\frac{C_\epsilon}{\overline C_\epsilon}\approx \left(1+\frac{\widetilde k}{\overline k}\right)^{-15/14}=\left(\frac{R_\lambda}{\overline R_\lambda}\right)^{-15/14}.
\end{equation}
and this is our prediction for the Reynolds number dependence of the normalized dissipation rate. 
To appreciate the similarity with the experimentally observed powerlaw (\ref{eq:CepsJCV}) one needs to realize that 
$\sqrt{R_L(0)}\sim \overline R_\lambda$ (combining expressions (\ref{eq:Ceps}) and (\ref{eq:Reynolds})) and that  $\overline C_\epsilon$ is a constant, so that this expression can be rewritten as
\begin{equation}\label{eq:Ceps1514JCV}
C_\epsilon\sim \left(\frac{\sqrt{R_L(0)}}{R_\lambda(t)}\right)^{15/14},
\end{equation}
and we find to a good approximation expression (\ref{eq:CepsJCV}). Indeed, the difference between  (\ref{eq:CepsJCV}) and (\ref{eq:Ceps1514JCV}) will in most cases be small enough to fall into experimental error-bars or the convergence of statistical averages in simulations. We further mention here also that in the experimental and numerical investigations reported in \cite{Vassilicos2015} the possiblity was left open that the exponents are not exactly, but only close to the ones in expression (\ref{eq:CepsJCV}).  

\begin{figure}
\begin{center}
\includegraphics[width=0.5\textwidth]{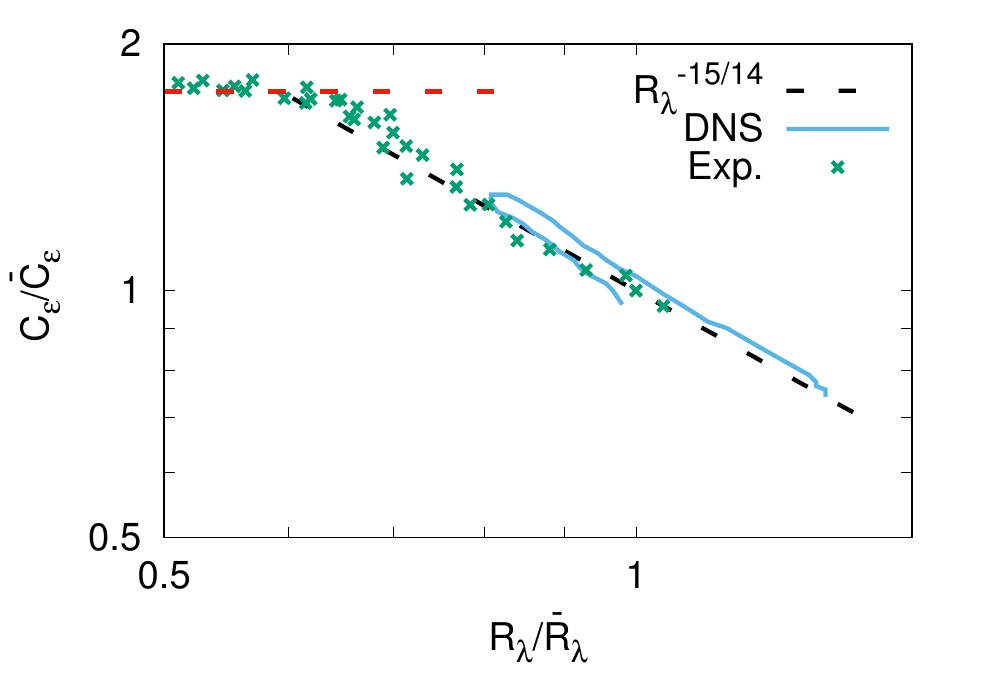}
\caption{The prediction of the Reynolds number dependence of the normalized dissipation rate, expressions (\ref{eq:Ceps1514}) and (\ref{eq:plnC}), compared to numerical \cite{Goto2015} and experimental \cite{Valente2012} results. \label{Fig:1}} 
\end{center}
\end{figure}

 At this point we will compare to existing results from literature. We first consider the case of a statistically steady state of isotropic, incompressible turbulence. In the simulations reported on in \cite{Goto2015} relation  (\ref{eq:K41}) was shown to be well satisfied for an important range of scales. If the flow is homogeneous, a space average will tend to the ensemble average if the volume over which is averaged contains  a sufficient number of flow structures. If this is not the case, temporal fluctuations will be observed around a long-time averaged flow.  This is the case in the simulations of reference \cite{Goto2015}. These box-averaged fluctuations are not necessarily in equilibrium and will thereby give rise to an evolution $C_\epsilon$. We have added to Figure \ref{Fig:1} the results of reference \cite{Goto2015} Figure 3,  for the fluctuations of $C_\epsilon$ around their average value for their highest Reynolds number. It is observed that those results are in perfect agreement with our prediction.

An obvious question is how these results apply to the case of a decaying turbulent flow where the new transient scaling was discovered. The comparison of our prediction with existing experimental results on grid-generated turbulence in a wind-tunnel is not straightforward since in the vicinity of the grid the turbulence is not statistically homogeneous, nor isotropic. We have however attempted a comparison with the experimental results reported in [\onlinecite{Valente2012}]. We have replotted in Fig. \ref{Fig:1} the data from their Fig. 6. We have considered their first data-point with a value of $C_\epsilon\approx 0.5$ to be the equilibrium state. It is observed that the experimental results, like the simulation, follow the theoretical prediction closely.

At this point, an open question is whether our analysis relevant for decaying turbulence at long times, where $C_\epsilon$ settles to a different constant value. We will consider the case where the kinetic energy decays following a powerlaw $k=k_0 (t/t_0)^{-n}$. The precise values of the reference quantities $k_0$ and $t_0$ are not important in the following. Deriving this expression for $k$ twice to obtain expressions for $\epsilon$ and $\dot \epsilon$ gives $\dot \epsilon/\epsilon=-(n+1)/t$ and $\epsilon/k=n/t$. Using these relations, integrating equations (\ref{eq:K41bar}) and (\ref{eq:Y94}) and eliminating $\kappa_0$ from the expressions, it is immediately found that 
\begin{equation}\label{eq:kkn}
\frac{\widetilde k}{k}=-\frac{2}{9}\frac{n+1}{n},
\end{equation}
and therefore, we find for the dissipation rate constant using (\ref{eq:Ceps1514}),
\begin{equation}\label{eq:plnC}
 \frac{C_\epsilon}{\overline C_\epsilon}\approx \left(\frac{9n}{7n-2}\right)^{15/14}.
\end{equation}
The value of $n$ is in general contained in the range $1\le n \le 2$, leading to the ratio $1.8\ge C_\epsilon/\overline C_\epsilon\ge 1.54$, which is a rather realistic range of values when compared to experiments and simulations \cite{Bos2007-2}. In Fig. \ref{Fig:1} we have added the asymptotic value ${C_\epsilon}/{\overline C_\epsilon}=1.75$ corresponding to a decay-exponent $n=1.2$, typical for decaying grid-generated turbulence, which fits the long-time data accurately.

These ideas explain why in the experimental and numerical results in \cite{Vassilicos2015} the Reynolds number decays before Taylor's expression is observed. Indeed, the imbalance is not a low-Reynolds number effect and in the experiments and simulations the Reynolds number is in principle high enough to observe both Taylor's and Kolmogorov's scaling. However, the evolution of both $R_\lambda$ and $C_\epsilon$ is a function of $\widetilde k/\overline k$. In turbulent flows in which the kinetic energy at long times decays following a power law, this latter quantity evolves from zero to a constant value, given by expression (\ref{eq:kkn}). The Reynolds number decays thus during the non-equilibrium transient from its initial value to a value $R_\lambda/\overline R_\lambda\approx (7n-2)/(9n)$.

We can now assess the reformulation of Taylor's expression (\ref{eq:Ceps}) for unsteady turbulence suggested in reference \cite{Goto2015}, given by 
\begin{equation}\label{eq:Deps}
 \epsilon= D_\epsilon (\overline U~ \overline L)\left(\frac{U}{L}\right)^{2}.
\end{equation}
Expression (\ref{eq:Ceps1514}) suggests the alternative scaling 
\begin{equation}\label{eq:2813}
 \epsilon= \overline C_\epsilon  \frac{U^3}{L} \left(\frac{\overline U^2}{U^2}\right)^{15/14},
\end{equation}
which shows that indeed the dissipation depends on the initial, equilibrium conditions, but, in contrast to (\ref{eq:Deps}) does not introduce any new constants. The constant  $\overline C_\epsilon$ is the equilibrium dissipation rate constant, which for our simplified inertial range spectrum is given by (\ref{eq:CepsBarNum}). Using the same arguments as above, we find that $D_\epsilon=\overline C_\epsilon \left({R_\lambda}/{\overline R_\lambda}\right)^{-1/7}$. The suggested quantity $D_\epsilon$ is therefore, according to our analysis, not strictly a constant, but its dependence on the ratio $R_\lambda/\overline R_\lambda$ is rather weak.

The present analysis is important for the modeling and understanding of turbulent flows since the non-equilibrium transient can be long and in many situations a self-similar decay might not even be reached before the flow is perturbed by the influence of boundaries, or because the Reynolds number has decayed too much for   (\ref{eq:K41}) and (\ref{eq:Ceps}) to be valid.  
Given the experimental evidence, the analytical results from the present letter suggest that the normalized dissipation in a wide class of unsteady turbulent flows can be described  
by the same, fairly simple, relation.

Expression (\ref{eq:Ceps1514}) constitutes the main result of the present letter. However, it is not the exact value of the exponent, which is close to the experimental observations, that is of interest. Indeed, its precise value can change slightly as a function of the detailed shape of the energy spectrum. We have checked this by assuming more realistic shapes for the energy containing range, and the results are robust, but the powerlaw exponent can somewhat change. 
What is of greater importance is that the foregoing analysis gives a firm theoretical basis for the transient behaviour of turbulent flows. The only non-trivial ingredient in the derivation is the shape of the unsteady energy-spectrum $E(\kappa,t)$ (expression (\ref{eq:Y94})). 
The present analysis complements thereby recent investigations suggesting that spectral imbalance \cite{Valente2014-2,Goto2016} and large scale coherence \cite{Goto2016-2} are behind the universal scaling of $C_\epsilon$ in non-equilibrium turbulence.
 
 Since this, rather simple, framework for unsteady turbulence allows to explain practically all the experimental observations in the transient, unsteady phase of developing turbulent flows \cite{Vassilicos2015}, it is plausible that engineering models can be improved by taking these ideas into account. We further think that the modeling and understanding of more complicated flows can greatly benefit from the insights obtained in this letter. For this to be successful, the ideas, here developed for isotropic turbulence, should be extended to other configurations such as shearflows and turbulent boundary layers. The simplicity of the present approach, using a low-frequency perturbative expansion of the equilibrium state, might make it applicable to these far more complicated situations.\\

The authors have benefited from discussion with J.C. Vassilicos.


\begin{thebibliography}{}

\end{thebibliography}


\begin{thebibliography}{10}

\bibitem{Leslie}
D.C. Leslie.
\newblock {\em Developments in the theory of turbulence}.
\newblock Oxford University Press, 1973.

\bibitem{FrischBook}
U.~Frisch.
\newblock {\em Turbulence, the legacy of A.N. Kolmogorov}.
\newblock Cambridge University Press, 1995.

\bibitem{KraichnanDIA}
R.H. Kraichnan.
\newblock The structure of isotropic turbulence at very high {R}eynolds
  numbers.
\newblock {\em J. Fluid Mech.}, 5:497--543, 1959.

\bibitem{Kraichnan65}
R.H. Kraichnan.
\newblock Lagrangian-history closure approximation for turbulence.
\newblock {\em Phys. Fluids}, 8:575, 1965.

\bibitem{Taylor1935}
G.I. Taylor.
\newblock Statistical theory of turbulence.
\newblock {\em Proc. Roy. Soc. London. Ser. A, Math. Phys. Sci.}, 151:421--444,
  1935.

\bibitem{Kolmogorov}
A.~N. Kolmogorov.
\newblock The local structure of turbulence in incompressible viscous fluid for
  very large {R}eynolds numbers.
\newblock {\em Dokl. Akad. Nauk. SSSR}, 30:301, 1941.

\bibitem{Sreeni84}
K.R. Sreenivasan.
\newblock On the scaling of the turbulence energy dissipation rate.
\newblock {\em Phys. Fluids}, 27:1048, 1984.

\bibitem{Kaneda2003}
Y.~Kaneda, T.~Ishihara, M.~Yokokawa, K.~Itakura, and A.~Uno.
\newblock Energy dissipation rate and energy spectrum in high resolution direct
  numerical simulations of turbulence in a periodic box.
\newblock {\em Phys. Fluids}, 15(L21), 2003.

\bibitem{Lumley92}
J.L. Lumley.
\newblock Some comments on turbulence.
\newblock {\em Phys. Fluids}, 4:206, 1992.

\bibitem{Bos2007-2}
W.~J.~T. Bos, L.~Shao, and J.-P. Bertoglio.
\newblock Spectral imbalance and the normalized dissipation rate of turbulence.
\newblock {\em Phys. Fluids}, 19:045101, 2007.

\bibitem{Seoud2007}
R.E. Seoud and J.C. Vassilicos.
\newblock Dissipation and decay of fractal-generated turbulence.
\newblock {\em Phys. Fluids}, 19(10):105108, 2007.

\bibitem{Mazellier2010}
N.~Mazellier and J.C. Vassilicos.
\newblock Turbulence without richardson--kolmogorov cascade.
\newblock {\em Phys. Fluids}, 22(7):075101, 2010.

\bibitem{Valente2012}
P.C. Valente and J.C. Vassilicos.
\newblock Universal dissipation scaling for nonequilibrium turbulence.
\newblock {\em Phys. Rev. Lett.}, 108(21):214503, 2012.

\bibitem{Valente2014}
P.C. Valente and J.C. Vassilicos.
\newblock The non-equilibrium region of grid-generated decaying turbulence.
\newblock {\em J. Fluid Mech.}, 744:5--37, 2014.

\bibitem{Discetti2013}
S.~Discetti, I.~B. Ziskin, T.~Astarita, R.J. Adrian, and K.~P Prestridge.
\newblock Piv measurements of anisotropy and inhomogeneity in decaying fractal
  generated turbulence.
\newblock {\em Fluid Dyn. Res.}, 45(6):061401, 2013.

\bibitem{Hearst2014}
R.~J. Hearst and P.~Lavoie.
\newblock Decay of turbulence generated by a square-fractal-element grid.
\newblock {\em J. Fluid Mech.}, 741:567--584, 2014.

\bibitem{Isaza2014}
J.~C. Isaza, R.~Salazar, and Z.~Warhaft.
\newblock On grid-generated turbulence in the near-and far field regions.
\newblock {\em J. Fluid Mech.}, 753:402--426, 2014.

\bibitem{Nagata2013}
K.~Nagata, Y.~Sakai, T.~Inaba, H.~Suzuki, O.~Terashima, and H.~Suzuki.
\newblock Turbulence structure and turbulence kinetic energy transport in
  multiscale/fractal-generated turbulence.
\newblock {\em Phys. Fluids}, 25(6):065102, 2013.

\bibitem{Nedic2013}
J~Nedi{\'c}, JC~Vassilicos, and B~Ganapathisubramani.
\newblock Axisymmetric turbulent wakes with new nonequilibrium similarity
  scalings.
\newblock {\em Phys. Rev. Lett.}, 111(14):144503, 2013.

\bibitem{Dairay2015}
T~Dairay, M~Obligado, and JC~Vassilicos.
\newblock Non-equilibrium scaling laws in axisymmetric turbulent wakes.
\newblock {\em J. Fluid Mech.}, 781:166--195, 2015.

\bibitem{Goto2015}
S.~Goto and J.C. Vassilicos.
\newblock Energy dissipation and flux laws for unsteady turbulence.
\newblock {\em Phys. Lett. A}, 379(16):1144--1148, 2015.

\bibitem{Yoshizawa1994}
A.~Yoshizawa.
\newblock Nonequilibrium effect of the turbulent-energy-production process on
  the inertial-range energy spectrum.
\newblock {\em Phys. Rev. E}, 49(5):4065, 1994.

\bibitem{Woodruff2006}
S.L. Woodruff and R.~Rubinstein.
\newblock Multiple-scale perturbation analysis of slowly evolving turbulence.
\newblock {\em J. Fluid Mech.}, 565:95, 2006.

\bibitem{Rubinstein2005}
R.~Rubinstein and T.T. Clark.
\newblock Self-similar turbulence evolution and the dissipation rate transport
  equation.
\newblock {\em Phys. Fluids}, 17(9):095104, 2005.

\bibitem{Horiuti2013}
K.~Horiuti and T.~Tamaki.
\newblock Nonequilibrium energy spectrum in the subgrid-scale one-equation
  model in large-eddy simulation.
\newblock {\em Phys. Fluids}, 25(12):125104, 2013.

\bibitem{Bos2007-3}
W.~J.~T. Bos, T.T. Clark, and R.~Rubinstein.
\newblock Small scale response and modeling of periodically forced turbulence.
\newblock {\em Phys. Fluids}, 19:055107, 2007.

\bibitem{Kovaznay}
L.S.G. Kovaznay.
\newblock Spectrum of locally isotropic turbulence.
\newblock {\em J. Aeronaut. Sci.}, 15:745, 1948.

\bibitem{Lumley1967}
J.L. Lumley.
\newblock Similarity and the turbulent energy spectrum.
\newblock {\em Phys. Fluids}, 10:855, 1967.

\bibitem{Goto2016}
S.~Goto and J.C. Vassilicos.
\newblock Local equilibrium hypothesis and {T}aylor's dissipation law.
\newblock {\em Fluid Dyn. Res.}, 48(2):021402, 2016.

\bibitem{Vassilicos2015}
J.C. Vassilicos.
\newblock Dissipation in turbulent flows.
\newblock {\em Ann. Rev. Fluid Mech.}, 47:95--114, 2015.

\bibitem{Valente2014-2}
P.C. Valente, R.~Onishi, and C.B. da~Silva.
\newblock Origin of the imbalance between energy cascade and dissipation in
  turbulence.
\newblock {\em Phys. Rev. E}, 90(2):023003, 2014.

\bibitem{Goto2016-2}
S.~Goto and J.C. Vassilicos.
\newblock Unsteady turbulence.
\newblock 2016.

\end{thebibliography}

\end{document}